\documentstyle[aps,twocolumn,epsfig]{revtex}

\begin{document}
\title{Electron transport through a metal-molecule-metal junction}
\author{C. Kergueris$^{a}$, J.-P. Bourgoin$^{a}$\thanks{%
Corresponding author. E-mail: jbourgoin@cea.fr}, S. Palacin$^{a}$, D. Esteve$%
^{b}$, C. Urbina$^{b}$, M. Magoga$^{c}$, C. Joachim$^{c}$.}
\address{a. Service de Chimie Mol\'{e}culaire,\\
b. Service de Physique de l'\'Etat Condens\'e, \\
CEA-Saclay \\
91191 Gif-sur-Yvette Cedex, France\\
c. CEMES-CNRS BP 4347, \\
31055 Toulouse Cedex, France}
\date{\bf{to Appear in Phys. Rev. B 59(19)1999}}
\maketitle

\begin{abstract}
Molecules of bisthiolterthiophene have been adsorbed on the two facing gold
electrodes of a mechanically controllable break junction in order to form
metal-molecule(s)-metal junctions. Current-voltage (I-V) characteristics
have been recorded at room temperature. Zero bias conductances were measured
in the 10-100 nS range and different kinds of non-linear I-V curves with
step-like features were reproducibly obtained. Switching between different
kinds of I-V curves could be induced by varying the distance between the two
metallic electrodes. The experimental results are discussed within the
framework of tunneling transport models explicitly taking into account the
discrete nature of the electronic spectrum of the molecule.
\end{abstract}

\newpage
\section{Introduction}

Molecular electronics, understood as `making an information processing
device with a single molecule' requires synthesizing molecules with
electronics functionnalities and connecting them together and to external
electrodes. In the seventies, A. Aviram and M. A. Ratner\cite{AVIRAM}, and
F. L. Carter\cite{CARTER} proposed challenging design of molecules analogous
to diodes or triodes. Since then, molecules of these types have been
synthesized, and some connection techniques have been developed. State of
the art is the connection of a few molecules, and even of a single one, to
conducting electrodes. This was first achieved using a Scanning Tunneling
Microscope by positioning a metallic tip above molecules deposited on a
conducting substrate\cite{AVIRAM.JOACHIM}. Electronic properties of various
molecules\cite
{LAMBIN,CYR,HAN,NEJOH,MIZUTANI,PORATH.MILLO,MICHEL,LU,BUMM,DEKKER,METZGER,DATTA.TIAN,DHIRANI}%
, and eventually single $C_{60}$ ones\cite
{JOACHIM.GIMZEWSKI.SCHLITTER,JOACHIMC60} have been investigated in this way.
The conclusion reached from these experiments is that a molecule can have a
non-zero conductance determined by its molecular orbital structure\cite
{JOACHIMC60,SAUTET,MUJICA}. Using a STM to contact a molecule however
suffers from intrinsic limitations, such as the asymmetry of the junctions
and, at least at room temperature, the lack of the mechanical stability
necessary to maintain a stable chemical bond between the molecule and the
tip. These limitations, together with the improvement of lithographic
techniques, have prompted the emergence of complementary techniques. In the
last three years, two alternative techniques have been proposed for
contacting a molecule to metallic electrodes. The first one consists in
fabricating a series of metallic electrodes, to which the molecule is
contacted, at the surface of a substrate. This technique, so far restricted
to molecules longer than at least 5 nm \cite{ROUSSET}, has been used, in
particular, to investigate carbon nanotubes\cite{TANS,EBBESEN}. The second
one is to use a mechanically controllable break ({\bf MCB}) junction\cite
{REED,MULLER}. It consists in breaking a small metallic wire, introducing
molecules with end-groups reactive to this metal into the gap, and adjusting
the gap between the two facing electrodes to a distance comparable to the
length of the molecule in order to contact it. This technique combines the
advantages of the previous ones: it allows to connect a short molecule
between similar electrodes while maintaining a very high stability (down to $%
0.2$ pm/hour for nanofabricated {\bf MCB }junctions)\cite{VANRUITENBEEK}. As
demonstrated recently by Reed et al\cite{REED}, it opens up the possibility
to measure electrical transport through a very few molecules chemically
bound to two electrodes, and possibly through a single one -the actual
number being however very difficult to know .

In the present paper, we describe the use of gold {\bf MCB} junctions to
investigate the electronic transport properties through bisthiolterthiophene%
{\bf \ }2.5''-bis(acetylthio)-5.2'5'.2'' -terthienyl ({\bf T3}) molecules.
We show (i) that different types of reproducible I-V characteristics can be
obtained and (ii) that varying the distance between the two electrodes
induces a switch between different I-V curves. Finally, we discuss the
interpretation of the experimental results within the framework of two
differents models which both explicitly include the discrete nature of the
electronic levels of the molecule: (i) a coherent model which treats the
molecule as a scattering impurity between two metallic wires, and (ii) a
sequential tunneling model, in which the molecule is assumed to be weakly
coupled through tunnel junctions to each metallic electrode.

\section{Experimental techniques and results}

\subsection{Sample fabrication}

\begin{figure}[b!]
\centerline{\epsfig{file=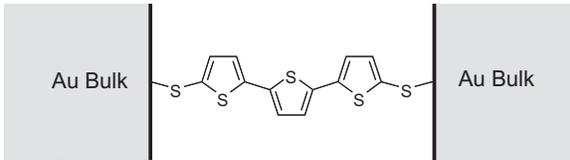,width=3in}}
\caption{Ideal Sample. A conjugated molecule is chemisorbed onto the gold
electrodes via the thiolate terminal groups.}
\label{FIG: 1}
\end{figure}

\begin{figure}[tbp]
\centerline{\epsfig{file=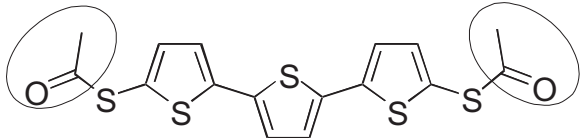,width=3in}}
\caption{Representation of the ${\bf T3}$ molecule. Acetyl protecting groups
are visible (circled) at each end. They were removed prior to the assembly
process.}
\label{FIG: 2}
\end{figure}

The aim of the experiment is sketched on Fig.\ref{FIG: 1}. The conjugated
molecule {\bf T3, }to be connected to the electrodes of the {\bf MCB}
junction, was synthesized from terthiophene. A thiolate function was
substituted at both ends of the molecule for its ability to strongly react
with gold surfaces\cite{ULMAN}. The thiolate functions were protected by
thioester formation with acetic anhydride in order to avoid successive
oxidative oligomerization that would generate polydisulfides \cite{TOUR} in
solution; the protecting acetate groups were removed just before immersion
of the gold electrodes into the {\bf T3 }solution (Fig\label{FIG: MOLECULE}%
\ref{FIG: 2}).

\begin{figure}[tbp]
\centerline{\epsfig{file=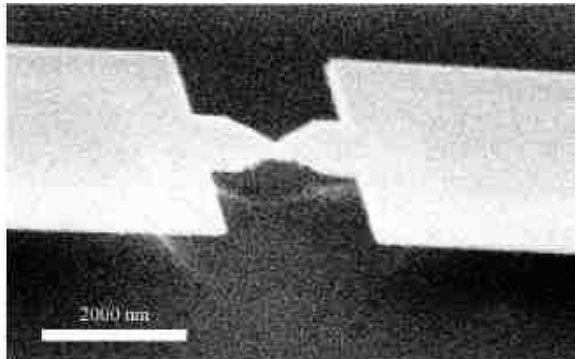,width=3in}}
\caption{Scanning electron microscope picture of a suspended junction before breaking.}
\label{FIG: 3}

\end{figure}

Suspended metallic microbridges were fabricated as described in Ref.\cite
{VANRUITENBEEK}. First, an insulating layer of polyimide PI 2610 from Dupont
de Nemours was spun on a polished phosphor-bronze substrate. Using standard
e-beam lithography techniques, a metallic nanostructure with the appropriate
shape was then deposited on a $4~{\rm mm}\times 20~{\rm mm}$ chip. In our
case, the metallic layer consisted in a 100 nm thick gold layer, with an
underneath 0.2 nm thick titanium adhesion layer and a 5 nm thick aluminum
protection layer on top. Finally, isotropic reactive ion etching of the PI
layer produced a metallic bridge suspended between anchoring pads (Fig.\ref
{FIG: 3}). The suspended length was of the order of 3 $\mu $m, and the
central constriction was less than 100~nm wide.

\begin{figure}[tbp]
\centerline{\epsfig{file=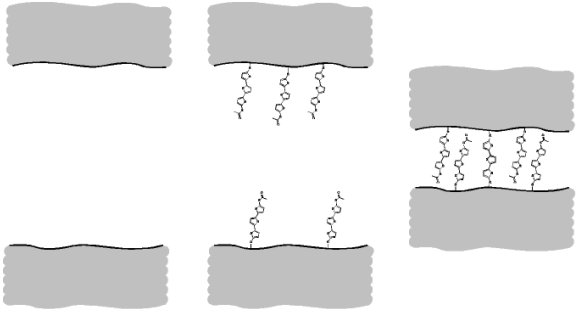,width=3in}}
\caption{Actual experiment: the top and bottom gold electrodes are first
separated by breaking the junction (left), {\bf T3 }molecules are adsorbed
onto them (middle) and the electrodes are brought closer to allow I-V
measurements (right) on a single or a small number of molecules.}
\label{FIG: 4}
\end{figure}

The chip was then mounted in a three point bending configuration. The chip,
resting on two countersupports, was bent by pushing on its center with a
driving rod, actuated through a coarse adjustment screw, until the bridge
breaks, as indicated by infinite electrical resistance. The molecules were
then self-assembled onto the freshly broken gold electrodes by immersion of
the suspended junction in a droplet of a $5\times 10^{-4}$ M solution of 
{\bf T3} in trichloro-1,1,1-ethane. The droplet was supplied by a Hamilton
syringe mounted 0.5 mm above the junction. In order to improve the
self-assembly process of the molecules, the acetyl protecting group of the
dithiol was hydrolized in situ by adding 0.1 $\%$ of dimethylaminoethanol 1
min before the experiment. The gold electrodes were kept in solution for 1
min, then the solvent was evaporated. After that, Argon was continously
flushed through the shielded sealed box containing the set-up. These
conditions were chosen so as to hinder the formation of di- or
polydi-sulfides. In a final step, the bridge gap was reduced using a
piezoelectric fine adjustment of the driving rod until a non zero conduction
was detected (Fig.~\ref{FIG: 4}).

\subsection{Experimental setup}
All electrical connections were filtered with lowpass RC filters. The
junction was voltage biased, and the current was measured using an I/V
converter. Typically, 512 points were collected for each I-V curve by
sweeping the voltage from -2 V to 2 V in 0.2 to 20 seconds.

\subsection{Control experiments and Calibration}

\begin{figure}[tbp]
\centerline{\epsfig{file=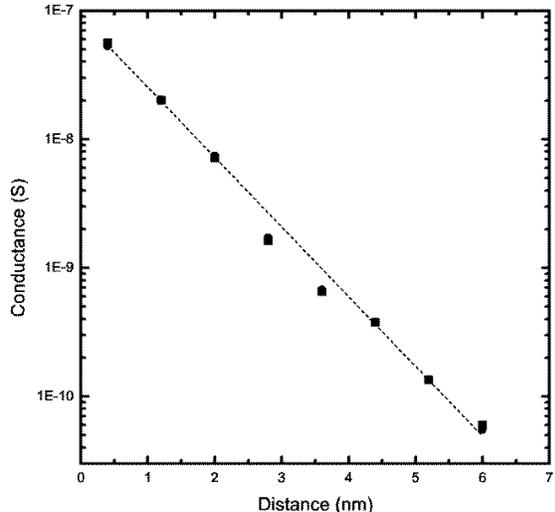,width=3in}}
\caption{Conductance of a metal-air-metal junction as a function of the
variation of the inter-electrode spacing (dots). The origin of the
horizontal axis is arbitrary. The dashed line correspond to the WKB
exponential variation assuming a barrier height of 1 eV(see text).}
\label{FIG: 5}
\end{figure}

As a control experiment, we first recorded the I-V characteristics of a
metal-air-metal junction. I-V characteristics are linear, and the variations
of the conductance with the piezoelectric actuator elongation are shown in
Fig.\ref{FIG: 5}. Assuming elastic deformation of the substrate and a
barrier height of 1 eV for a gold-air-gold tunnel junction\cite{LEBRETON},
the observed exponential dependence\cite{SIMMONS} provides a calibration of
the displacement ratio $r$ between the piezo elongation and the
inter-electrode spacing. The obtained value $r=3.3\times 10^{-5}$ is
consistent with the geometrical estimate \cite{VANRUITENBEEK,KEIJSERS} $%
r=6tu/L^{2}=3.75\times 10^{-5}$, where $t=0.3$ mm is the substrate
thickness, $u=3$ $\mu $m is the distance between the anchoring points, and $%
L=1.2$ cm is the distance between the two countersupports.

I-V characteristics of a metal-air-metal junction after 1 min immersion in
the pure solvent were also measured. I-V curves wer featureless and
exhibited a linear behaviour at low bias.

\subsection{Measurements of Au-{\bf T3}-Au junctions}

\begin{figure}[tbp]
\centerline{\epsfig{file=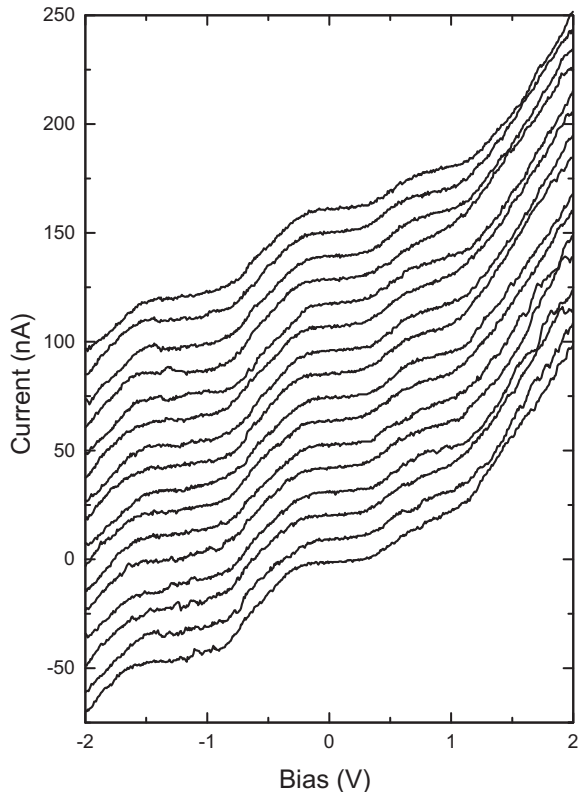,width=3in}}
\caption{Consecutive single sweep I-V curves recorded at room temperature
for a gold-{\bf T3}-gold junction. Curves are shifted vertically for
clarity. }
\label{FIG: 6}
\end{figure}

\begin{figure}[tbp]
\centerline{\epsfig{file=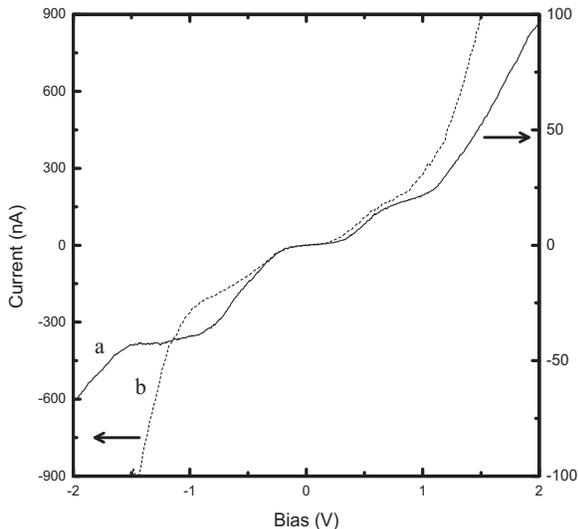,width=3in}}
\caption{Typical {\bf (a)} asymmetric (solid line) and {\bf (b)} symmetric
(dashed line) I-V curves recorded at room temperature for gold-{\bf T3}-gold
junctions. Both curves were obtained by averaging over 5 voltage sweeps.}
\label{FIG: 7}
\end{figure}

During a typical experiment on Au-{\bf T3}-Au junctions, stability periods
with a duration 1-20 min alternate with instability periods generally
lasting a few minutes. This behavior was always observed on the four samples
that have been measured. A series of I-V characteristics, recorded
subsequently, are shown in Fig.\ref{FIG: 6}. Although different I-V
characteristics could be observed, the reproducible ones are of one of the
two types of I-Vs shown in Fig.\ref{FIG: 7}. Asymmetric I-Vs of type {\bf (a)%
} are more often observed and more stable than symmetric ones of type {\bf %
(b)}.

\begin{figure}[tbp]
\centerline{\epsfig{file=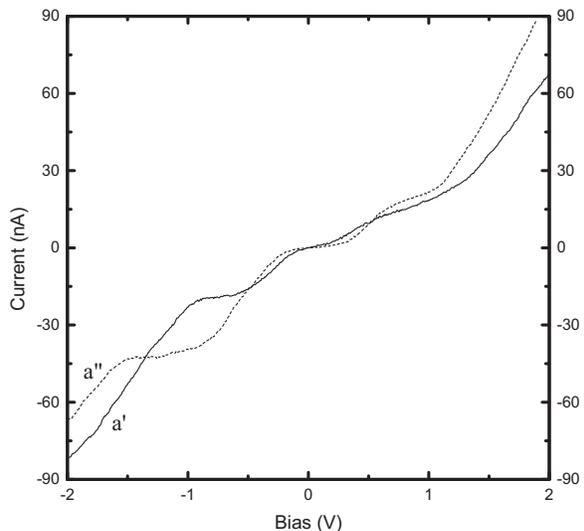,width=3in}}
\caption{I-V curves recorded (a') before (solid line) and (a'') after
(dashed line) reduction of the inter-electrode spacing by ca 0.4nm. Both I-V
curves were obtained by averaging over 5 voltage sweeps.}
\label{FIG: 8}
\end{figure}

Fig.\ref{FIG: 8} shows typical asymmetric curves.

The measured zero bias conductance of type {\bf (a) }junctions is of the
order of 10 nS. {\bf \ }The asymmetric I-V characteristic is non-linear with
step-like features, and the current increases linearly at large voltage.

The measured zero bias conductance of type {\bf (b)} junctions is larger,
about 80 nS. The symmetric I-V characteristic is also non-linear with
smaller step-like features. At $V\geq 1$ V, the current rises faster than
linearly with $V$.

Fig.\ref{FIG: 8} shows a mechanically induced transition between different
I-V curves. A series of reproducible type {\bf (a') }I-Vs was first
recorded. The inter-electrode spacing was then reduced by 0.04 nm
approximately, and a\ series of type {\bf (a'')} I-Vs were then recorded. It
should be noted that upon reduction of the gap size, the zero bias
conductance first {\it decreased} from 13 nS to 6 nS.

\section{Discussion}

Before discussing transport models, we first present the electronic
properties of the isolated {\bf T3} molecule and their modification upon
adsorption onto the electrodes.

\subsection{Electronic properties of a {\bf T3} molecule}

\begin{figure}[tbp]
\centerline{\epsfig{file=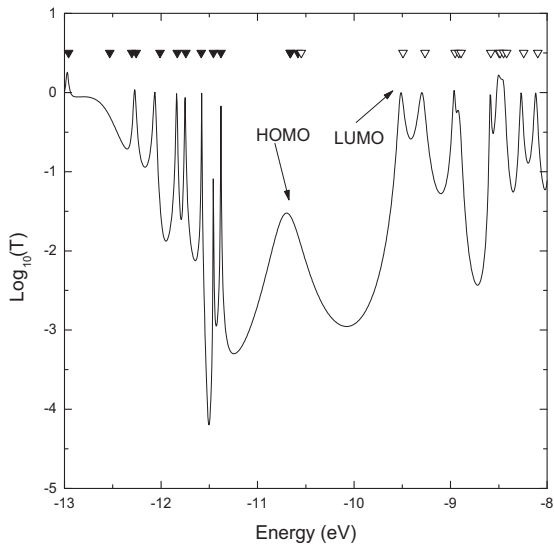,width=3in}}
\caption{ Transmission function of the ${\bf T3}$ molecule calculated by the
 ${\bf ESQC}$ technique. The triangles represent the energy level positions of
the isolated molecule ( filled occupied MO's and open unoccupied MO's of the
deprotected ${\bf T3}$ diradical). HOMO and LUMO refer to the gap of 
${\bf T3}$
adsorbed on the two electrodes. The energy scale reference is arbitrary.}
\label{FIG: 9}
\end{figure}

Fig.~\ref{FIG: 9} (triangles) shows the electronic spectrum of an isolated
deprotected {\bf T3} molecule calculated using a standard extended
H\"{u}ckel technique, starting from a forced planar geometry optimized at
the AM1 level\cite{PARAMEH}. Although properties of the {\bf T3} molecule
adsorbed onto gold are not known, except for the adsorption of {\bf T3} in a
matrix of dodecanethiol \cite{NANOTECH}, other rigid rods $\alpha ,\omega $%
-dithiols have been shown to form assemblies onto gold, in which one thiol
group binds to the surface while the other thiol group projects upward at
the outer surface of the self-assembled monolayer\cite{TOUR}. Furthermore,
organised self-assembled monolayers of
bis-(2,2':5',2''-terthien-5-yl)disulfide have been described \cite{LIEDBERG}%
. From these studies, it seems plausible that the {\bf T3} molecule, once
deprotected, reacts onto the gold electrodes with (i) the formation of a
Au-S bond and (ii) the terthiophene moiety pointing upward rather than
laying on the surface of the electrode. One key question concerns the
position of the Fermi level of the electrodes relatively to the molecular
electronic levels. The latter are expected to shift so that the Fermi level
of gold falls in the HOMO-LUMO gap, its position being determined by the
amount of charge displaced on the molecule. It has been shown that the
formation of the Au-S bond involves a formal negative charge transfer , $%
\delta ^{-}$ from the metal to the molecule. The values of $\delta ^{-}$
ranges from 0.2 to 0.6 electron for thiolate adsorption on gold\cite
{DATTA.TIAN,DHIRANI,SELLERS}\cite{NOTE1}. Using the method described in
reference\cite{TIAN}, we found a position of the Fermi Level of $-10.5\pm
0.06$ eV in the energy scale used for the molecule. The HOMO is thus closer
to the Fermi level than the LUMO, with $E_{F}-E_{HOMO}=0.2$ $\pm 0.06$ eV%
\cite{NOTE2}. However, given the approximations made in the calculation and
the uncertainty about the exact amount of charge transferred, it seems
reasonable to keep this difference as an adjustable parameter within the $%
[0\ -0.7$eV$]$ range, i.e. Fermi Level between HOMO and midgap.

\subsection{Linear regime}

At low bias ($V\leq 0.1$V), the I-V characteristics are linear. The measured
zero bias conductance are respectively 80nS, 6nS and 13nS for the curves in
Fig.\ref{FIG: 8} and Fig.\ref{FIG: 7}.

We first calculated the transport properties of the gold-molecule-gold
junction by the Electron Scattering Quantum Chemistry ( {\bf ESQC) }technique%
\cite{SAUTET,JOACHIM.VINUESA,MAGOGA}. This technique has proven to
quantitatively account for the low bias conductance of single adsorbed
molecules as probed by STM\cite{JOACHIM.GIMZEWSKI.SCHLITTER}. It treats the
molecule as a defect which breaks the translational invariance of the metal,
and therefore scatters incident electrons. The {\bf ESQC} technique ignores
both the electron-electron and electron-phonon interactions and neglects
charging effects. It assumes that the scattering is elastic because for
molecule of small length, the tunneling time is shorter than the
intramolecular relaxation times. An extended H\"{u}ckel model is used to
build up the matrix representation of the multichannel scattering
Hamiltonian taking into account the complete chemical description of the
electrodes and of the molecule. The calculated multichannel transmission
coefficient $T(E)$ of an electron at a given energy $E$ is shown in Fig.~\ref
{FIG: 9}. The linear conductance $G$ of the metal-molecule-metal junction is
then determined using the Landauer formula\cite{BUTTIKER},

\[
G=\frac{2e^{2}}{h}T(E_{F}) 
\]

where $E_{f}$ is the Fermi level of the electrodes.

The prediction for $G$ thus depends on the estimated position of the
electronic spectrum relatively to the Fermi energy $E_{F}$, on the exact
conformation of the molecule in the junction and of the coupling of the
molecule to the electrodes. The strength of coupling is determined by the
length of the Au-S bonds. The S atoms were assumed to be adsorbed in a
hollow site of the gold surface. A bond length of $1.905$ \AA\ was used in
the present calculation\cite{SELLERS}. This is the shortest distance, we
found in the literature. It thus provides an upper bound for the coupling
strength. Assuming a symmetric coupling at both ends of the molecule, the
calculated conductances for $E_{F}-E_{HOMO}=0.6$(midgap), $0.2$ and $0$ eV
are $87,585,2306$ nS, respectively. Although the order of magnitude of these
values is comparable with the measured one $G\simeq $80~nS for type ({\bf b}%
) junctions \cite{NOTE4}, the discrepancy indicates that the coupling of the
molecule to the electrodes is smaller than estimated. This finding can be
compared with the results published recently by Emberly and Kirczenow\cite
{EMBERLY}. In that paper, it is shown that the experimental conductance of
gold-benzene di -thiol-gold junctions, reported in ref. \cite{REED} can be
accounted for by a transport model based on the Landauer formalism provided
that artificially large bond lengths (i.e. small couplings) are used. In our
case the discrepancy between theory and experience seems less important but
could also be reduced using longer Au-S bonds: for example G$_{0}$=87nS at
midgap for Au-S=1.905\AA\ reduces to G$_{0}$=20nS at midgap for Au-S=3\AA .

\subsection{Non linear regime}

We now discuss the step-like features found in the I-V characteristics at
large bias voltage and their possible interpretations.

At first, we can rule out the simplified Coulomb blockade model put forward
by Reed et al to account for their experimental observations in a similar
experiment with benzene dithiolate \cite{REED}. Indeed, this model, which
considers the molecule as a usual metallic island forming a small capacitor
with each electrode, predicts a series of steps which are not observed.

Consequently, we attribute the steps to the discreteness of the molecular
levels\cite{NOTE3}. We present and discuss below two models which both
involve the discrete electronic levels of the molecule explicitly. The first
one is a coherent tunneling model derived from the {\bf ESQC} technique,
whereas the second is a sequential tunneling model. Basically, they differ
by the fact that the electron resides (sequential tunneling) or not
(coherent tunneling) on the molecule during the transfer.

Given the symmetry of the molecule, symmetric I-V characteristics are
expected in the case of symmetric couplings. Thus type ({\bf b})
characteristics only will be analyzed below.

\subsubsection{Coherent tunneling model}

The coherent tunneling model is based on an extension of the {\bf ESQC}
technique outlined above. It supposes that the coupling to the electrodes is
strong, in other terms that the tunneling time of an electron through the
molecule is much smaller than the intramolecular vibronic relaxation time:
i.e. the molecule is not charged by the tunneling process, the electron
having no time to be completely relocalized in the molecule after the
initial tunneling step at the metal -molecule electronic contact. The
current is calculated using \cite{DATTA}: 
\begin{equation}
I(V)=\frac{e^{2}}{\pi \hbar }\int_{-\infty }^{+\infty }T\left( E,V\right)
\left( f\left( E-\mu _{1}\right) -f\left( E-\mu _{2}\right) \right) dE
\label{EQ:landauer}
\end{equation}

in which the effect of the bias voltage $V$ on the $T(E)$ spectrum is
included; $\mu _{1,2}$ is the chemical potential of the electrode 1,2, and ${%
f(\epsilon )}$ denotes the Fermi function at the temperature of the
experiment. In the case of a symmetric junction, the voltage at the molecule
lies halfway between the electrode voltages. In order to calculate the
current, we make the crude approximation that $T(E,V)\simeq T(E-eV/2,0)$
where $T(E,0)$ is the spectrum of Fig. \ref{FIG: 9}. Equation \ref
{EQ:landauer} then predicts the I-V characteristic shown in Fig.~\ref{FIG:
10}.

\begin{figure}[tbp]
\centerline{\epsfig{file=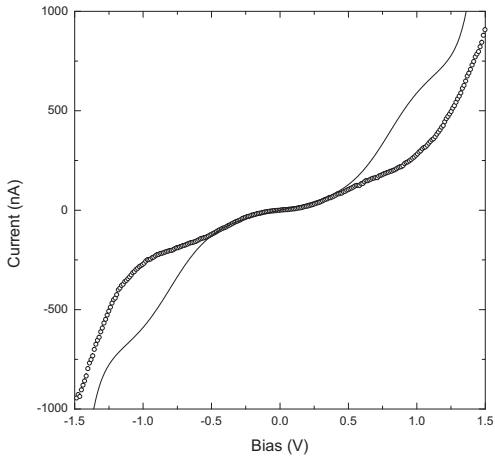,width=3in}}
\caption{Experimental symmetric I-V curve (diamonds). I-V curve (solid line)
calculated with the coherent model. The calculation (see text) has been done
using: $E_{F}-E_{Homo}$=0.4eV.}
\label{FIG: 10}
\end{figure}
 For both polarities, the first resonances occur through the HOMO and
the HOMO-1 levels. This results from the assumption that the equilibrium
Fermi level is closer to the HOMO than to the LUMO \cite{DATTA.TIAN}. In
this model, the shape and height of the steps is only due to the broadening
of the molecular levels by the coupling with the electrodes. In particular,
the height of a step in the I-V curve is directly proportional to the area
of the corresponding peak in the T(E,V) spectrum. The experimental curve is
well reproduced for the $[-0.5V,0.5V]$ range. Outside of this range, the
calculated current is higher than the experimental one. It should be noted
that the only adjustable parameter used in the calculation is the position
of the Fermi level. Here we have taken $E_{F}-E_{HOMO}=0.4$ eV. In this
calculation, the magnitude of the current depends on the following points:
i) the overlap between the orbitals of the outer Au atoms and of the sulfur
atoms; ii) the localisation of the molecular orbital and its symmetry, iii)
perfect symmetry between the two contacts\cite{KERGUERIS}. The applied
voltage is expected , especially at high bias, to affect points ii) and
iii), both of which should tend to decrease the calculated current and thus
reduce the discrepancy.

\subsubsection{Sequential tunneling model}

In this model, the molecule is treated as a quantum dot with discrete energy
levels weakly coupled to both electrodes through tunnel junctions. It
considers the tunneling of an electron through the metal-molecule-metal
junction as a sequential process\cite{PORATH.LEVI}, the molecule being
successively charged and discharged. At first sight, the existence of tunnel
barriers at both ends of the molecule can be questionned since the sulfur
species chemically adsorbed on the electrodes do contribute to the HOMO and
LUMO, both orbitals being delocalised over the entire molecule. However, the
existence of barriers to the injection of carriers at metal-oligothiophene
junction have been reported in the literature\cite{ZIEGLER}. Furthermore,
the possible existence of a barrier to injection at a gold
-thioacetylbiphenyl interface has also been recently put forward by Zhou et
al\cite{ZHOU} to interpret experimental data of conduction in a
metal/self-assembled monolayer/metal heterostructure.

Electron tunneling rates through both metal-molecule junctions are obtained
from a Golden rule calculation. The current is calculated by solving a
master equation connecting the different charge states of the molecule (see
Appendix A).

To calculate the I-V characteristics, we have taken the apparent position of
the molecular levels given by the $T(E)$ spectrum of Fig.~\ref{FIG: 9} in
order to account for the shift of the level upon adsorption of the molecule
onto the electrodes. For the sake of simplicity, we adjusted the
experimental curve considering only two levels (HOMO and LUMO) because the
charging of the molecule will most likely involve these two levels. The best
adjustment of the experimental data is obtained assuming the HOMO to be
located at the equilibrium Fermi level\cite{NOTE0}. For this calculation, a
charging energy of $E_{c}$=$0.19$ eV was used, which is comparable to values
reported for C$_{60}$ molecules by Porath et al \cite{PORATH.LEVI}. This is
about one order of magnitude smaller than the calculated charging energy for
an isolated molecule but a strong reduction of the charging energy is
expected for molecules adsorbed onto metallic electrodes.

\begin{figure}[tbp]
\centerline{\epsfig{file=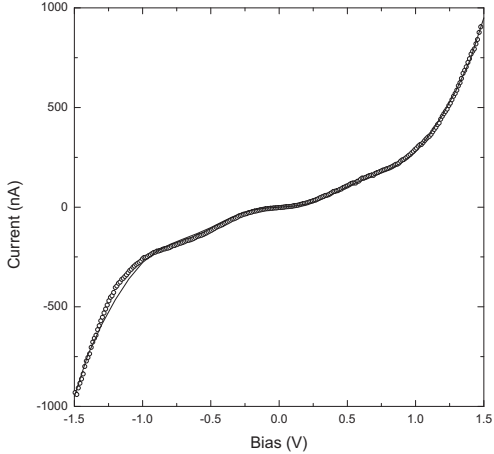,width=3in}}
\caption{Experimental symmetric I-V curve (diamonds). I-V curve (solid line)
calculated within the framework of the sequential model. The calculation
(see text) has been done using: $E_{F}=E_{Homo}$, $\phi $= 0.47 eV and $d$ =
0.53 nm .}
\label{FIG: 11}
\end{figure}

As can be seen in Fig.~\ref{FIG: 11}, a good qualitative description of the
experimental data can be obtained once the tunneling barrier parameters are
adjusted to fit the magnitude of the current step. A prefactor is included
to account for the barrier suppression at high bias\cite{KOROTKOV} (See
Appendix~A). The phenomenological parameters (accounting for the barrier
height and width respectively) used to fit the data are $\phi $ =0.47eV and $%
d$=5.3\AA . This approximation is a very crude way to account for the
contact barriers at both ends of the molecule. However, though the value for
the barrier width is large, the orders of magnitude for $\phi $ and $d$ are
not irrealistic.

A more microscopic approach for the transparencies of the barriers would be
to use the extended Huckel calculation to compute the overlap of the
molecular orbitals (involved in the change of the total number of electrons
in the molecule) with the Au leads and then estimate the rates from the
broadening of these levels upon coupling. In the present case, the levels
involved in the transport are the HOMOs levels. We have calculated the
broadening of these levels using the ESQC technique assuming a Au-S distance
of 1.905 \AA\ and considering each molecular orbital separately to avoid any
interference effect between orbitals (such an interference effect takes
place, for example, between the HOMO and HOMO-1 orbitals in the T(E)
spectrum in Fig. \ref{FIG: 9} and explains the non unity value of the
corresponding peak). We found a width at half maximum of 0.4eV corresponding
to a transfer rate of about 3 10$^{14}$\ Hz. This is too high when compared
to the experimental rates of the order of 10$^{12}$\ Hz. To get the right
order of magnitude artificially large Au-S distances have to be used
(typically 6 \AA ). This shows that the calculation of the coupling strength
based on the extended H\"{u}ckel model likely leads to an overestimated
value as already pointed out in the linear regime subsection above. At
present, this microscopic approach unfortunately cannot be used to estimate
the barrier suppression at high bias, because the extended H\"{u}ckel
calculation do not take into account the modification of the orbitals under
bias.

\subsubsection{Discussion}

To summarize, both the coherent model at low bias and the sequential model
at large bias give qualitative fits of the experimental I-V curves. This
provides support to the hypothesis that the step like features in the I-V
curve indeed originate in the discreteness of the molecular levels. However
a quantitative fit of the data has not been achieved. This may be explained
as follows. First, the effect of the applied bias on the relative position
of the resonances in the $T(E)$ spectrum is neglected, since we consider
that the bias effect is to shift $T(E)$ as a whole. Although the electric
fields involved in the experiments are always lower than $1\times 10^{9}$
V/m, the position of the levels is expected to be modified \cite{LAZZARONI}.
Preliminary calculation using the DFT formalism have shown that the
HOMO-LUMO gap of the bisthiol-bisthiophene molecule is indeed reduced under
an electric field \cite{BUREAU.KERGUERIS}, and this effect is expected to
take place also with {\bf T3}. Another effect of the field is to modify the
coupling between the molecule and the electrodes. This should show up in the 
$T(E,V)$ spectrum as a modification of the width of the peaks and
consequently in the I(V) as a modification of the height of the steps.

Second, in the calculation, the molecule is assumed to be in the vacuum. In
reality, given the experimental conditions, the molecule can be assumed to
be surrounded by Argon and probably a few solvent molecules. We have not
taken this environment into account in our calculations. Smoothing of the
theoretical I-Vs of both models is expected due to this polarizable
surrounding.

Additional experimental results and theoretical calculations are clearly
needed to help deciding between sequential and coherent tunneling or a
combination of both. This includes : (i) measurements in a better controlled
environment (UHV), and at low temperature to determine the origin of the
broadening of the steps in the I-V curve, (ii) calculation of the potential
profile along the molecule in order to determine wether tunneling barriers
are present at the end of the molecule or not and (iii) estimation of the
tunneling times involved in the processes. Furthermore, it should be noted
that two proposed model indeed correspond to extreme cases. In the coherent
one, the charging effect is neglected, whereas in the sequential one, the
quantum coherence is neglected. In our qualitative fits, it turns out that
the order of magnitude of the molecular level broadening (coherent model)
and charging energy(sequential model) are comparable. Thus a proper model
will very likely have to take into account both coherent and sequential
transport.

Compared to the symmetric case, the interpretation of the asymmetric I(V)
curves of Fig.~\ref{FIG: 7} (curve a) and Fig \ref{FIG: 8} is more
complicated because we can no longer assume that the coupling to the
electrodes is symmetric at both ends of the molecule. First, these data
likely correspond to the case where at least one end of the molecule is less
coupled to the electrode than in the case of the symmetric curve since i)
the zero bias conductance is smaller than in the symmetric case and ii) the
high bias behaviour of the current is linear-like rather than
exponential-like. Second, the effect of mechanically pushing on the
molecule(s) induces a change between two different stable curves. Such an
effect has been used recently at low bias to realise an electromechanical
amplifier based on a C60 molecule\cite{JOACHIM.GIMZEWSKI}. In the present
case, two explanations of the observations can be put forward: either the
coupling of the molecule to the electrodes is changed by the mechanical
action, or a change in the conformation of the molecule was induced.

\section{conclusion}

In this paper, we have investigated the transport properties of molecules of
2,5''-bis(acetylthio)-5,2',5',2''-terthienyl self-assembled in the
adjustable gap of a metallic break junction. We have observed that the I-V
characteristics recorded at room temperature are not always symmetric with
respect to the polarity of the applied bias and show two different regimes:
a linear regime at low bias $V<0.1$ V and a highly non-linear regime with
step-like features at higher voltage. The order of magnitude of the measured
zero bias conductance is comparable to the theoretical calculation made with
the {\bf ESQC} technique assuming a single molecule in the gap of the break
junction. This indicates that these experiments likely involve a very few
molecules. We have obtained qualitative fits of the symmetric I-V curves
using two different models of transport, a coherent one for the low bias
range and a sequential one for the high bias range, both of which explicitly
involve the electronic structure of the molecule. Finally, we have shown
that the mechanical action of decreasing the gap size by $0.04$ nm\ induces
a strong modification of the I-V curves.

{\bf Acknowledgments :} We thank C. Bureau and M. Devoret for enlightening
discussions, C.\ Blond and T.\ Risler for their contribution to the
sequential model calculations, O. Araspin for his help with the e-beam
lithography.\newpage

\appendix

\section{\label{app:A}The Sequential Model}

\subsection{Description of the system}

\begin{figure}[tbp]
\centerline{\epsfig{file=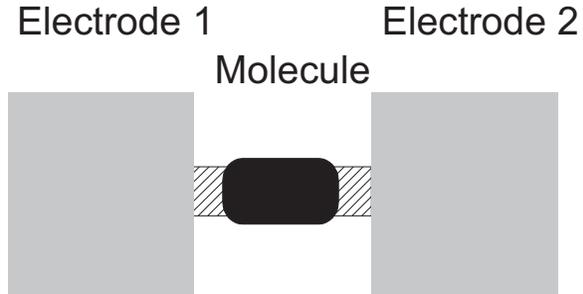,width=3in}}
\caption{Scheme of the metal-molecule-metal junction, consisting of a
molecule weakly coupled to electrodes via tunnel barriers (hatched)}
\label{FIG: 12}
\end{figure}

We consider a molecule weakly coupled via tunnel barriers to two electrons
reservoirs (Fig.\ref{FIG: 12}). We assume that the tranfer of an electron
through the metal-molecule-metal junction occurs through subsequent
tunneling events at both ends of the molecule

The reservoirs are described by a continuum of states ( temperature $T$ and
Fermi energy $E_{F}$, occupation according to the Fermi-Dirac statistics $%
f\left( E-E_{F}\right) =\left( 1+\exp \left( \left( E-E_{F}\right)
/kT\right) \right) ^{-1}$). Under bias $V$ the electrochemical potentials of
the electrodes are given by 
\begin{equation}
\left\{ 
\begin{array}{l}
\mu _{2}=E_{F}+\left( 1-\eta \right) eV \\ 
\mu _{1}=E_{F}-\eta eV
\end{array}
\right.  \label{_Energie électrodes}
\end{equation}
where $\eta $ is a parameter used to describe the strength of coupling at
both ends of the molecule. In the case of the symmetric coupling, we have $%
\eta $=1/2.

The molecule is accounted for by a set of $K$ discrete molecular orbitals.
The $i^{th}$ state of the molecule is described by its energy $E_{i}$ and
the vector $\left( \lambda _{1}^{i},\lambda _{2}^{i},\ldots ,\lambda
_{K}^{i}\right) \,$where $\lambda _{k}^{i}$ is the occupation number of the $%
k^{th}$ molecular orbital of energy $\varepsilon _{k}$. The corresponding
total number of electrons in the molecule is $n_{i}=\sum_{k=1}^{K}\lambda
_{k}^{i}$. The energy of the $i^{th}$ state is 
\begin{equation}
E_{i}=\sum_{k=1}^{K}\lambda _{k}^{i}\varepsilon _{k}+E_{C}\left(
n_{i}-n_{g}\right) ^{2}  \label{_Energie de l'état i}
\end{equation}

where $E_{C}$ is the Coulomb energy and $n_{g}$ is the ground state number
of electrons in the molecule plus the offset charge.

\subsection{Calculation of the transfer rates}

The rate of transfer of electrons through the tunnel barriers are calculated
by application of the Fermi golden rule assuming that the tunneling
processes are elastic.

The rate $\Gamma _{i,j}^{k+}$ for the process connecting the $i^{th}$ state
and $j^{th}$ state through the transfer of one electron from electrode k
onto the molecule is given by. 
\[
\Gamma _{i,j}^{k+}=\frac{2\pi }{\hbar }\left| T_{i,j}\right| {^{2}\rho }%
_{k}f\left( E_{j}-E_{i}\right) 
\]

where $T_{ij}$ is the tunnel matrix element coupling the states $i$ and $j$,
and ${\rho }_{k}$ is the density of states of the electrode k, assumed
constant around the Fermi level.

We define in the same way $\Gamma _{i,j}^{k-}$ connecting the states $i$ and 
$j$ by transferring one electron from the molecule onto electrode k. 
\[
\Gamma _{i,j}^{k-}=\frac{2\pi }{\hbar }\left| T_{i,j}\right| {^{2}\rho }%
_{k}\left( 1-f\left( E_{i}-E_{j}\right) \right) 
\]

Assuming that the tunneling barriers have a height $\phi $ and a width $d$,
we apply a WKB type approximation for the tunneling matrix element 
\[
\left| T_{i,j}\left( V_{b}\right) \right| {^{2}}\text{ }\alpha \text{ }\exp
(-2\frac{\sqrt{2m(\phi -eV_{b})}}{\hbar }d) 
\]
where $V_{b}$ is the voltage drop across a tunnel junction.

\subsection{Calculation of the current}

The matrix $\underline{\underline{\Gamma }}$ connecting the different states
of the molecule is \cite{Weinmann1996} 
\begin{equation}
\underline{\underline{\Gamma }}=\underline{\underline{\Gamma }}^{1+}+%
\underline{\underline{\Gamma }}^{1-}+\underline{\underline{\Gamma }}^{2+}+%
\underline{\underline{\Gamma }}^{2-}  \label{_Taux total}
\end{equation}

We introduce the vector $\underline{P}$ , which $i^{th}$ component is the
probability of having the molecule in state $i$. The evolution of the system
is then described by 
\begin{equation}
\frac{d}{dt}P_{i}=\sum_{j}\left( \Gamma _{ji}P_{j}-\Gamma _{ij}P_{i}\right)
\label{_Equation maitresse}
\end{equation}

defining $\underline{\underline{M}}$ by $M_{ij=}\Gamma _{ji}$ $i\neq j$ and $%
M_{ii=}-\sum_{j}\Gamma _{ij}$ equation \ref{_Equation maitresse} becomes, 
\[
\frac{d}{dt}\underline{P}=\underline{\underline{M}}\underline{P} 
\]

We look for a stationary solution $\underline{\widehat{P}}$ assuming that a
permanent regime rapidly sets in 
\[
\underline{\widehat{P}}=\ker \underline{\underline{M}} 
\]

The current is finally obtained by calculating the net number of electrons
transferred from right to left through either junction: 
\begin{equation}
I=e\sum_{i,j}\widehat{P}_{j}\left( \Gamma _{ji}^{1-}-\Gamma
_{ji}^{1+}\right) =-e\sum_{i,j}\widehat{P}_{j}\left( \Gamma
_{ji}^{2-}-\Gamma _{ji}^{2+}\right)  \label{_courant}
\end{equation}


\begin{references}
\bibitem{NOTE2}  {\sl The use of a shift in energy of the molecular levels
with regard to the Fermi level of the electrodes is a formal convenient way
to account for a more complex reality. Indeed, the adjustment of the
molecular levels to the electrodes levels should be accounted for in the
density of states of the metal electrodes, a term we consider here as a
constant and presently hidden in the calculation. }
\bibitem{NOTE0}  {\sl A temperature of 600 K was used in the calculation to
account for a possible broadening of the level due to the adsorption or
induced by the surrounding.}
\end{references}
\end{document}